# Detecting quench in HTS magnets with LTS wires — a theoretical and numerical analysis

Rui Kang, Juan Wang, Qingjin Xu

Key Laboratory of Particle Acceleration Physics & Technology, Institute of High Energy Physics, Chinese Academy of Sciences, Beijing 100049, China

E-mail: xuqj@ihep.ac.cn; kangrui@ihep.ac.cn

**Abstract**

Protecting a high temperature superconducting (HTS) magnet from a quench event is a challenging task. Because of the slow normal zone propagation velocity, the detection method by directly monitoring coil voltage may not be timely for HTS anymore, leaving HTS magnets under danger of overheating. Using a NbTi low temperature superconducting (LTS) wire to detect quench in coils wound with ReBCO HTS tapes have recently been experimentally proved, yet a theoretical study is still needed to further develop this technique and make it prepared to be applied more generally in high field magnets. In this manuscript, we have demonstrated that it is the significant difference in the temperature dependence of critical current between LTS and HTS but not the normal zone propagation velocity (NZPV), that makes LTSs good quench detectors. Simulations show that LTS quench detectors should have low matrix fraction or high matrix resistivity. At last, at field up to 15 T or 20 T, $Nb_3Sn$ is proven to be a good quench detector.

## I. Introduction

HTS materials, represented by the ReBCO coated conductor, appear advantages over LTS to be used in magnets generating a magnetic field beyond 15 T, like in controlled nuclear fusion reactors and high energy accelerators [1,2]. However, protecting an HTS magnet from a quench event, which might happen due to a local disturbance or defect, is still a challenging task. As the name indicates, HTS has higher critical temperature ($T_c$) than LTS, and thus the critical current ($I_c$) of HTS decreases with temperature much slower than that of LTS, as is shown in Figure 1. The scaling relations for NbTi, $Nb_3Sn$, $Nb_3Al$, $MgB_2$ and ReBCO to plot figure 1 are respectively taken from [3–7]. The absolute critical current of ReBCO, NbTi and $Nb_3Sn$ are scaled to the state-of-art commercial products [8,9]. These relations are also used in the later analysis. Once a normal zone is established, it propagates much slower in HTS than in LTS. Consequently, the long reliable quench detection method by coil voltage may not be timely for HTS anymore. A later quench detection is however fatal to a superconducting magnet, as the large current will keep generating heat at the normal zone. This issue is especially critical in accelerator magnets where conductor current density is generally very high.

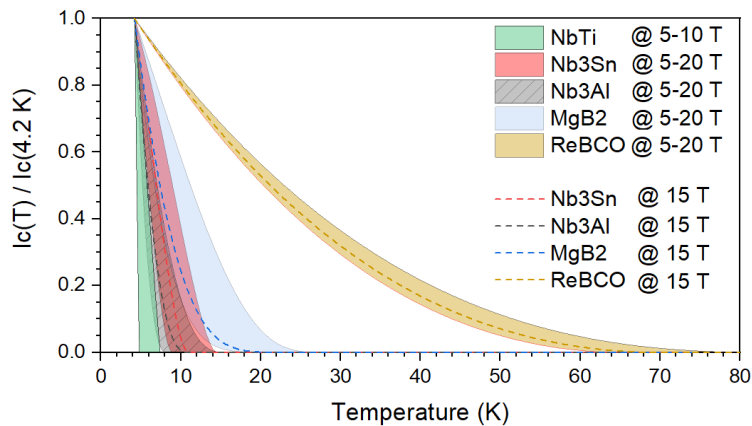

Figure 1. The variation of critical current as function of temperature and magnetic field for ReBCO and several LTSs. The lower field limit is corresponding to the right boundary. For ReBCO and three candidates for high field quench detectors, the curves at 15 T are drawn with dash lines.

Non-Insulation winding technology is a promising approach to solve the quench protection problem for HTS, but it doesn't suit all magnets. As another route, many different quench detection methods are being developed for HTS, such as by optic fibers [10–12], acoustic signals [13], radio-frequency wave [14], stray-capacitance [15] or even fluorescent thermal imaging[16]. Despite that these methods are not ready to replace quench detection by voltage yet, some of them are only applicable to special magnets because of the physical quantities they measure. Apart from the above-mentioned technologies, a brilliant idea to detect quench in HTS magnets was proposed by Hasegawa et al [17–19], by using another superconductor. In one example, an insulated NbTi wire was attached on a ReBCO pancake coil, and a small monitoring current (~3 A) was applied to the NbTi wire. When the ReBCO pancake coil went to quench, a sharp voltage transition from neglectable to more than 1 V was observed in the NbTi wire, at which time voltage in the ReBCO coil was still below 10 mV. Such an approach doesn't reply on any special medium and is in principle feasible to any kind of superconducting magnet. Similar technique was also suggested more recently by Gao et al [20], for faster detecting quench in a $Nb_3Sn$ CCT dipole magnet, with a co-wound $Nb_3Sn$ wire.

In this manuscript, a more detailed explanation of why and how a LTS wire can work as a quench detector for HTS is studied. The purpose is generally to answer these questions:
1) What is the principle that allows LTS wires to detect quench in HTS magnets?
2) What kind of LTS wire can best work as a quench detector?
3) Can LTS quench detector be more efficient than detecting quench in HTS directly by its voltage?

As a start, a theoretical discussion is presented in section II by analyzing the characteristics of several superconductors. Then section III presents the simulated results on a stacked ReBCO cable, together with a discussion about what kind of LTS wire can best work as a quench detector for HTS. At last, in Section IV, the application of this technique on the high field, where HTS is destined to play a role, is discussed. Several candidates of high field LTS quench detectors are compared.

## II. A theoretical explanation

Concerning quench performance, a general idea about the difference between HTS and LTS is their normal zone propagation velocity (NZPV). However, here it will be demonstrated it is not the NZPV, but the resistance-temperature characteristic that makes LTS a possible sensor for a quench event in HTS.

For practical application, the voltage-current characterization of a superconductor is usually described by the E-J power law [21]. In each superconductor, the superconducting composition can be regarded equipotential with the normal metal matrix [22], and the total current in the two kinds of components should equal to the charged current, which gives:

$$E_c \left(\frac{I_{sc}}{I_c}\right)^n = \eta_n \frac{I_s - I_{sc}}{A_n} \quad (1)$$

In the above equation, $E_c$ stands for a critical electrical field, regarded as the threshold whether the conductor is in superconducting state. n is the index value. $I_s$, $I_{sc}$ and $I_c$ are respectively the total current charged to the conductor, current in the superconducting component and critical current of the conductor. $\eta_n$ is the resistivity of the metal component, and $A_n$ is the cross section area of normal metals. The proper $I_{sc}$ that meets the above equation can be find by an iteration method [22], which then gives the right value of electrical field and joule heating. A MATLAB code is developed to do the numerical calculation. Although for HTS and LTS, different $E_c$ may be selected (1 or 0.1 μV/cm, the former is selected here) and their n index may vary (10-40 for HTS and 30-80 for LTS, see page 370 in [23]. Here 40 is chosen in all cases.), the most important difference comes from their $I_c$. As is already presented in Figure 1, the temperature dependence of $I_c$ for HTS and LTS can be very different. As a result, the voltage-temperature behavior can also be completely different.

Figure 2 shows the Resistance (per unit length) versus temperature curves for three typical superconductors. The specifications of these wires are listed in Table I. For NbTi, a very sharp voltage transition happens only few kelvins above the operating temperature, depending on the background magnetic field. This sharp transition comes from losing of superconductivity. Then its resistance continues to grow because of increase of resistivity of copper with temperature. Note the magneto resistance of copper also plays a role. For $Nb_3Sn$ wires, the transition happens at higher temperature due to its high $T_c$. As for ReBCO, voltage doesn't show a sudden transition, but grows up gradually when temperature exceeds the current sharing temperature ($T_{cs}$). Such disparity of Resistance-temperature characteristic between HTS and LTS gives an opportunity to detect the quench of HTS by the latter, which doesn't reply on propagation of normal zone, as will be further demonstrated by a numerical analysis in the next section.

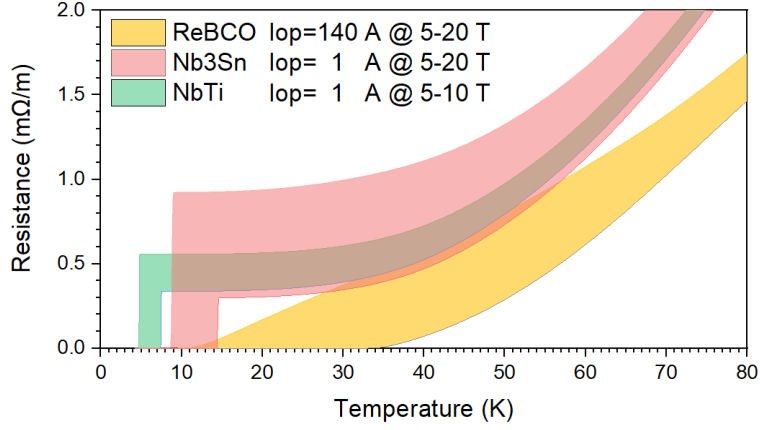

Figure 2. The variation of resistance (per unit length) as function of temperature and magnetic field for ReBCO, Nb$_3$Sn and NbTi. The operating current (I$_{op}$) of ReBCO takes example of the case in section IV. A small monitoring current of 1 A is assumed for both LTSs. The lower field limit is corresponding to the right boundary.

Table I. Wire specifications corresponding to Figure 2.

| Wire type | Materials ratio | Geometry | RRR of copper |
|---|---|---|---|
| ReBCO | ReBCO: Cu: Hastelloy = 1:40:50 | 4 mm × 0.091 mm | 100 |
| NbTi | NbTi: Cu = 1:1 | Φ 0.8 mm | 100 |
| Nb$_3$Sn | Nb$_3$Sn: Cu = 1:1 | Φ 0.8 mm | 150 |

## III. Quench simulation of a ReBCO tape stack attached with a NbTi wire

### 1. Simulation model

As an example, this section numerically analyzes the thermal and electrical behavior of a ReBCO tape stack and an attached NbTi wire initially operated at 4.2 K and a background field of 5 T. A thicker copper is assumed in the ReBCO tape to keep a moderate copper current density of 500 A/mm². The related parameters are listed in Table II.

Table II. Specifications of ReBCO cable and NbTi quench detectors for quench simulation implemented in this section.

| Wire Type | Materials ratio | Geometry | Copper RRR | Ic at 4.2 K and 5 T [A] | Calculated Tcs with Iop [K] |
|---|---|---|---|---|---|
| ReBCO Cable | ReBCO: Cu: Hastelloy = 1:180:50 | 4 mm × 0.231 mm ×10 | 100 | 4518 | ~10.9 |
| NbTi detector #1 | NbTi: Cu = 1:1 | Φ 0.8 mm | 100 | 616 | ~7.40 |
| NbTi detector #2 | NbTi: Cu = 1.78:1 | Φ 0.25 mm | 100 | 77 | ~7.36 |

Numerical analysis of an adiabatic superconducting system with two kinds of superconductors insulated with each other can be achieved by solving the below 1-D heat balance equations with finite difference method in MATLAB when the longitudinal length is much longer than its sizes in cross section [22]:

$$A_i \rho_i C p_i \frac{\partial T_i}{\partial t} - \frac{\partial}{\partial x}\left(A_i k_i \frac{\partial T_i}{\partial x}\right) + \frac{(T_i - T_j)}{H_m} = \dot{q}'_{ext} + \dot{q}'_{Joule} \quad (2)$$

The subscript i and j indicate ReBCO cable or the LTS detector. For each conductor, A is the cross-section area, ρ is the average density, C$_p$ is the average specific heat capacity, k is the average thermal conductivity. The average thermal properties of each wire are calculated according to the law of mixture. The third term indicates the heat transfer between the two components, where H$_m$ is their thermal contact resistance (TCR). The two terms in right are respectively applied heating (to initiate quench in a simulation) and joule heating, which is calculated by solving eq. (1). For two insulated superconductors at DC condition, there is no electrical coupling between them.

A conductor length of 2 m is imagined, which is long enough to avoid boundary effect and requires moderate meshes. Them mesh size is 4 mm and time step is 10 μs. The TCR between the ReBCO stack and the NbTi is initially assumed 0.1 K·m/W, which stands for a good heat conduction at least at low temperatures. This crucial parameter will be discussed later. Heat pulses with different power, distributing over 0.01 m and lasting for 0.01 s, are applied on the center of the HTS stack. The thermal properties of the materials refer the data base of the THEA code [24]. Results show the minimum quench energy (MQE, expressed here in power per unit length) is about 600 W/m.

## 2. Results and discussion

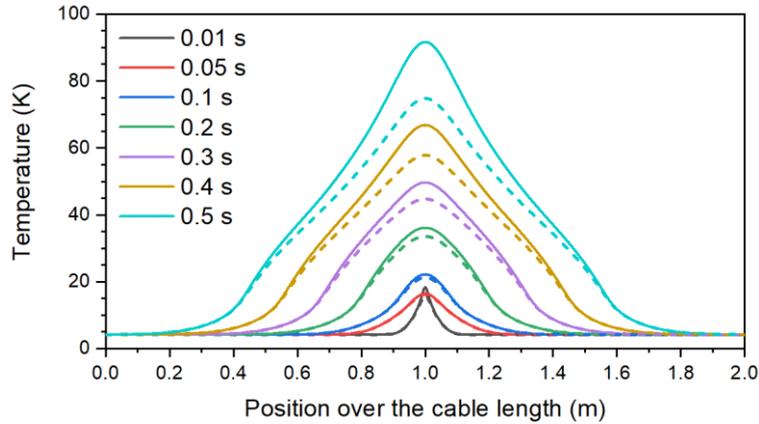

Figure 3. Temperature profiles of the ReBCO stack (solid lines) and attached NbTi wire (dashed lines) at different times. The frontier of temperature profiles for both conductor is totally the same, indicating no faster normal zone propagation in the NbTi wire.

Figure 3 shows the evolution of temperature profile over the two conductors with 600 W/m power applied. At a first glance, it is clear that normal zone, or more precisely the frontier of high temperature profile, is not going to propagate faster in the LTS quench detector than in the HTS cable. It is because that the current charged to the LTS detector is so small, that it generates so little heat and can hardly drive the normal zone to propagate. In addition, the ReBCO cable is working as a huge heat sink. Temperature in the LTS detectors increase only because of the heat transfer from ReBCO Cable. Therefore, if a LTS can work as quench detector for HTS, it is definitely not because of its higher NZPV.

Because of the rather high copper current density of the ReBCO stack, the hot spot temperature increases dramatically to above 100 K in 500 ms, while the normal zone, if defined by temperature above the $T_{cs}$ temperature, only propagates for about 1 m. Note such NZPV is already a rather high value for ReBCO, because of the high current, low operating temperature and large copper content imagined for this example [25]. Figure 4 compares the voltage, hot spot temperature and normal zone length of the two conductors, all shown in dual log axis to see details at early times. Another recovery case is also compared. The voltages of ReBCO stack with the two deposited energy respectively show the typical thermal run away and recovery behavior after the external heat pulse switched off at 10 ms, at which time the voltage is still quite low. In practice, a voltage threshold around 100 mV is usually used to determine a magnet is quenching because of the inevitable noises. According to this value, the quench of the HTS stack can be recognized after 300 ms, despite that a delay time is still needed for validation and switching on the protection system.

The voltage over the LTS wire presented in figure 4(a), on one hand, do start to increase in less than 1 ms when the temperature reaches its $T_{cs}$, much earlier than HTS. From figure 4(c), the normal zone length is still at the level of centimeters. This further helps to conclude that it is not fast NPZV make LTS a good quench detector for HTS. Note that the normal zone length in NbTi detector is still longer than that in the ReBCO cable, which is only because of the lower $T_{cs}$ of the former. On the other hand, in this example the voltage in the LTS wire is quite small (at the level of $10^2$ μV). Besides, it is difficult to judge if the HTS stack is going to quench or recovery only by voltage of the LTS wire until after 100 ms. This example seems not demonstrate very well the idea to detection HTS quench with a LTS wire.

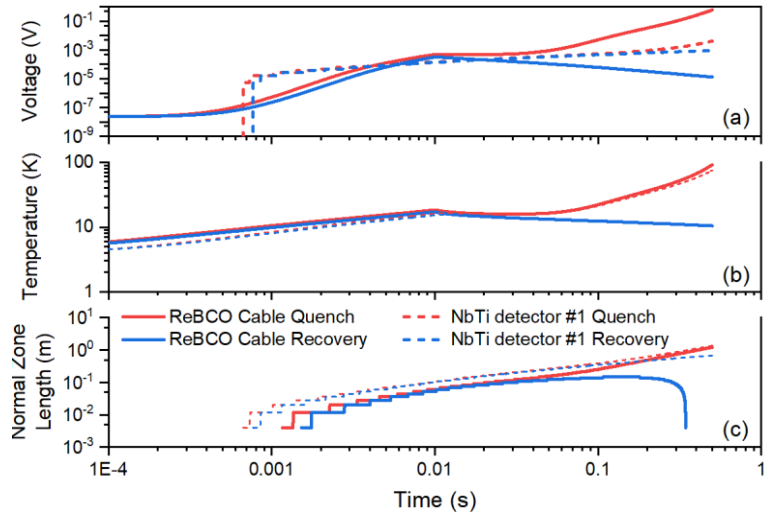

Figure 4. Variation of voltage, hot spot temperature and normal zone in ReBCO conductor and NbTi detector as function of time. Quench and recovery cases are compared.

The small voltage in LTS sensor is due to the small current in the LTS wire and the rather large copper cross section. In a LTS quench detector, while the small current is obligatory, the copper cross section is redundant. The 0.8 mm NbTi wire imagined in this case contains half fraction of copper, which is the usual case for winding magnet. As a result, even all current goes to copper, the electrical field is still only about 13 μV/cm at 4.2 K and 5 T. It should be noted that for the usual LTS wire, a considerable amount of pure copper is mandatory for stability or quench protection, because the operating current is high. However, when it works as a detector with a very small working current, both of them would not be problems. As a consequence, if a LTS wire is designed to be a HTS quench detector, it should have a small diameter, with a small fraction of copper or with other matrix having higher resistivity, or even get the matrix removed, as was done by Hasegawa. The ideal one would be the superconducting filament itself. Figure 5 shows the voltage in the LTS detector if the NbTi is replaced by a 0.25 mm one (marked as NbTi detector #2) with thickness of copper layer 0.02 mm (so the ratio between copper and superconductor is about 0.56). The voltage in this thin LTS wire significantly increased.

As is mentioned above, the voltage threshold for validating a quench in a magnet is usually set around 100 mV because of the noise signals, which may come from the inductive contribution or the ripple of power supply itself during charging the magnet. While the inductive noises can be much released by techniques like monitoring the voltage difference of two similar coil sections, the one from power supply is not easy to get rid of. On the other hand, the LTS quench detector should be powered a constant small current, so the noises from power supply is also expected to be very small. Even if the magnet generates some inductive voltages in LTS detector, it can be easily cancelled. Therefore, the voltage threshold for detecting quench with LTS wires can in principle be set at a rather low value, as long as it is high enough to be distinguished from a stability event.

## 3. Effect of TCR between HTS cable and LTS quench detector

In the early simulations, the TCR is assumed a rather low value. However, this critical parameter may vary over a wide range due to different insulation layers, contact conditions and impregnation or not. As a reference, as Hasegawa et al reported, when a Φ0.3 mm NbTi is attached to ReBCO tape with epoxy, the thermal resistance is around 0.02 Km²/W. When accounting the wire diameter, it is equivalent to a TCR around 67 K·m/W. A poor thermal contact condition could be deathful for such quench detection method. In figure 5, the voltage evolution in the NbTi detector #2 with different TCR are compared. It can be noted that, as TCR increases, the presence of sharp voltage transition does delay, but with modest values. Even if the TCR is increased to a large value of 100 Km/W, NbTi detector can still present a sharp voltage at around 15 ms when its temperature exceeds the $T_{cs}$, much earlier than the HTS cable reaching 100 mV (about 300 ms). Note that with the high copper current density (500 A/mm²), the hot spot temperature in HTS cable is dramatically increasing. Therefore, LTS quench detector still shows superiority even if it has a relatively poor heat transfer with the ReBCO cable.

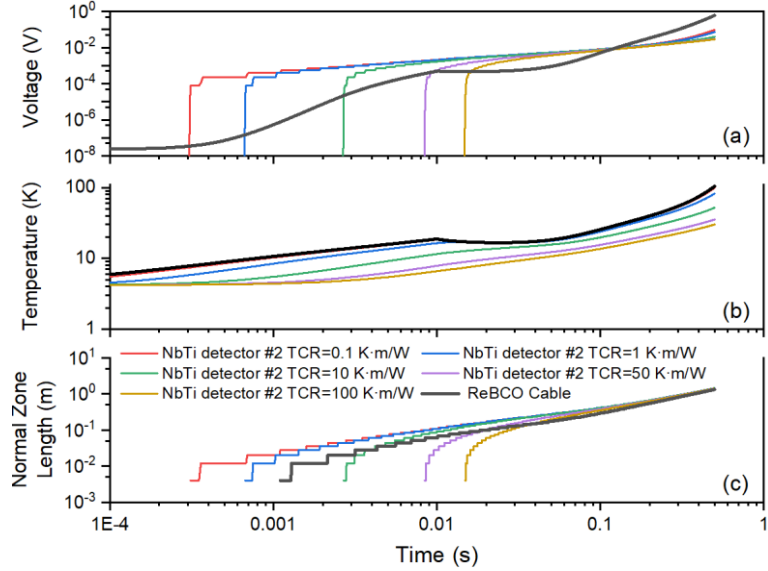

Figure 5. Variation of voltage, hot spot temperature and normal zone in ReBCO conductor and NbTi detector as function of time, with different TCRs.

## 4. High field application

In practice, ReBCO is most likely to be applied at magnetic field higher than 15 T, which is beyond the capability of NbTi wires. Apart from it, other LTS materials like $Nb_3Sn$, $Nb_3Al$ and $MgB_2$ are also not commonly considered to build magnets at this field region, because of their low $I_c$. However, they may eventually contribute themselves well as quench detectors. In this section, the feasibility of these materials working as quench detectors for a ReBCO cable operating at 4.2 K and 15 T is studied by simulation. The parameters of ReBCO cable refers a prototype of a novel Roebel structure cable under developing at our lab for high field accelerator magnets, as shown in Table III. Based on the scaling relation mentioned above, the critical current of the cable at 4.2 K and 15 T is about 15.9 kA. Taking $I_{op}/I_c$ as 80%, the $T_{cs}$ of the cable is then about 10 K. As a start point, the quench detector should have some critical current at 4.2 K, and loss superconductivity at around 10 K. Coincidently, the above mentioned three LTS all show such behavior. Table III lists the configuration of LTS wires for this analysis. $Nb_3Sn$ refers a standard product from [4,8]. $Nb_3Al$ and $MgB_2$ are much less commercialized than $Nb_3Sn$ or NbTi. As a result, two wires are taken for examples from publications [5,6]. Although at 4.2 K and 15 T, the critical current of these three kinds LTS can be significantly different, all of them loss their superconductivity at temperature around 10 K with a slight difference, as earlier shown in figure 1. As discussed above, the resistivity of matrix is another crucial condition for the detection capability. Despite the three kinds of wires may be produced with quiet different matrix, here they are all assumed the same matrix as $Nb_3Sn$ wire (pure copper), but the original filling factors are kept. In such a way, the different detection capability would mainly represent the superconducting properties of these superconductors themselves. A good TCR of 0.1 Km/W is assumed for the simulations in this section. Due to a lack of thermal properties of $MgB_2$ and $Nb_3Al$ at different temperatures, the specific heat capacity and thermal conductivity of these two superconductors are also assumed the same as $Nb_3Sn$. Since the cross-section area of ReBCO cable is much larger than these LTS detectors, and a low TCR is assumed, such assumption shall not change the result.

Table III. Specifications of ReBCO cable and LTS quench detectors relevant to 15 T application.

| Wire Type | Materials ratio | Geometry | RRR of copper | Ic at 4.2 K and 15 T [A] | Calculated Tcs with Iop [K] |
|---|---|---|---|---|---|
| ReBCO Cable | ReBCO: Cu: Hastelloy = 1:40:50 | 4 mm × 0.091 mm ×72 | 100 | 15900 | ~10.2 |
| $Nb_3Sn$ detector | $Nb_3Sn$: Cu = 1:1 | Φ 0.818 mm | 150 | 122.6 | ~10.6 |
| $Nb_3Al$ detector | $Nb_3Al$: Cu = 1:1.5 | Φ 0.507 mm | 150 | 133.4 | ~9.9 |
| $MgB_2$ detector | $MgB_2$: Cu = 1:6.63 | Φ 0.83 mm | 150 | 4.4 | ~10.2 |

Figure 6 shows the simulated results. All three LTS detectors show sharp voltage transition when hot spot temperature of ReBCO cable is around 10 K, but with a slight difference. Sharp voltage transition first shows in $Nb_3Al$ wire at about 4 ms when hot spot temperature at ReBCO cable reaches 10 K. Then after few milliseconds voltage transition occurs in both $Nb_3Sn$ and $MgB_2$ wires, but in $MgB_2$ wire the transition is a little bit relaxing. These behaviors also agree with their $I_{op}/I_c$-T relations. Later, when temperature further increases, three curves show complete the same trend, which comes from the temperature dependent resistivity of copper. It should be noted that although sharp voltage occurs in all three LTS detectors, the absolute values are all at the level of 1e-4 V, even less than the voltage signals in ReBCO cable. This is because of the high fraction of high purity copper matrix are assumed for all of them. Once the matrix is replaced with metals or alloys with higher resistivity or the amount is reduced (like using acid), the voltage signals in LTS detectors can be dramatically increased. As is discussed above, loaded with a small current, these LTS quench detectors don't need large amount of low resistive matrix to ensure stability or quench protection.

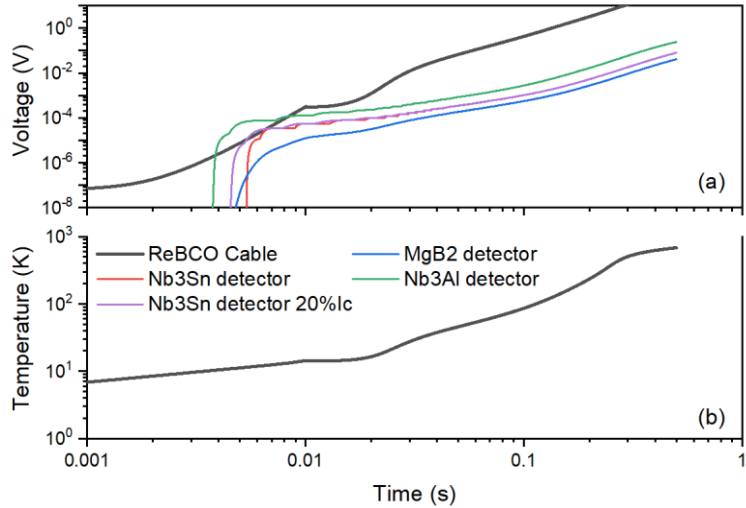

Figure 6. Variation of voltage (a) and hot spot temperature (b) in ReBCO conductor and the attached LTS candidates for high field application. Good thermal contact is assumed.

This example takes an operating field at 15 T, and the $T_{cs}$ of ReBCO cable is happened to be around 10 K. From figure 1, we can see that even at 20 T, these three LTS wires can still carry some superconducting but loss it at a little bit higher temperature. They don't make big differences when working as quench detector. In this sense, $Nb_3Sn$ would be the best option as quench detectors for high field ReBCO magnets, since it is commercialized.

Unfortunately, there is a last hurdle $Nb_3Sn$ must step over, which is its strain sensitivity. When applied in ReBCO magnets, the $Nb_3Sn$ quench detector must be already been through the heat treatment, which means the brittle $Nb_3Sn$ wire could be easily degraded during the magnet assembly. Therefore, it is important to figure out if a $Nb_3Sn$ can still work if its critical current is greatly reduced. Figure 6 also shows the voltage signal in the $Nb_3Sn$ wire with the same condition as the above case, but the critical current is reduced to 20%, which is corresponding to a tension of about 0.65% or a compression of about -0.7% at 4.2 K and 15 T according to the scaling law reported in [4]. Quench of the ReBCO cable could still be recognized, even earlier. The absolute value of $I_c$ of the LTS quench detectors is not as crucial as when it is used to build a magnet. Actually, the $MgB_2$ case also sets such an example: the $MgB_2$ wire chosen for the above discussion has a small critical current of 4.4 A at the imagined condition, but it also work well as quench detector.

## 5. Summary

The significant difference in temperature dependence of critical current between LTS and HTS, makes LTSs good quench detectors for HTS, and the quench detectivity does not rely much on the normal zone propagation.

The LTS quench detector will be loaded with a small monitoring current, thus is free from stability problem or quench protection. High purity copper matrix should be removed as much as possible or be replaced with high resistive materials to enhance the detectivity of LTS wires.

As such quench detection strongly relies on the heat transfer from HTS cable to LTS detector, their thermal contact resistance (TCR) would have important influence on the quench detectivity. Even though, simulation shows that with a rather poor TCR of about 100 K·m/W, a NbTi wire can still detector a quench event in 20 ms, much earlier than voltage in HTS cable reaching a reasonable detectable threshold (about 300 ms in the simulated case to reach 100 mV).

When ReBCO is applied to high field up to 20 T, Nb$_3$Sn, Nb$_3$Al and MgB$_2$ could all work as quench detector. As the most commercialized candidate, Nb$_3$Sn seems to be the most promising LTS quench detector at high field. It has also been proven by simulation that the absolute critical current of the LTS quench detector is rarely relevant, so some degradation due to strain could be allowed during assembling, so long the wire is not totally damaged. Note this doesn't mean Nb$_3$Al or MgB$_2$ have no chances. Comparing to Nb$_3$Sn, Nb$_3$Al can tolerate much higher strain, and MgB$_2$ has high critical temperature, so they may become irreplaceable quench detection in some cases.

Bases on these results, we'll later test the feasibility of Nb$_3$Sn quench detector on ReBCO insert coils. Besides, a proper voltage threshold or a strategy that can efficiently recognize a quench event will also be studied by both experiment and simulation.